

\input epsf.tex

\font\rmu=cmr10 scaled\magstephalf
\font\bfu=cmbx10 scaled\magstephalf

\font\it=cmti10 scaled \magstephalf

\rmu

\font\rmus=cmr8
\font\rmuss=cmr6
\font\mait=cmmi10 scaled\magstephalf
\font\maits=cmmi7 scaled\magstephalf
\font\maitss=cmmi7
\font\msyb=cmsy10 scaled\magstephalf
\font\msybs=cmsy8 scaled\magstephalf
\font\msybss=cmsy7
\font\bfus=cmbx7 scaled\magstephalf
\font\bfuss=cmbx7
\font\cmeq=cmex10 scaled\magstephalf

\textfont0=\rmu
\scriptfont0=\rmus
\scriptscriptfont0=\rmuss

\textfont1=\mait
\scriptfont1=\maits
\scriptscriptfont1=\maitss

\textfont2=\msyb
\scriptfont2=\msybs
\scriptscriptfont2=\msybss

\textfont3=\cmeq
\scriptfont3=\cmeq
\scriptscriptfont3=\cmeq

\newfam\bmufam  \textfont\bmufam=\bfu
      \scriptfont\bmufam=\bfus \scriptscriptfont\bmufam=\bfuss

\hsize=15.5cm
\vsize=21cm
\baselineskip=16pt   
\parskip=12pt plus  2pt minus 2pt

\def\a{\alpha}
\def\b{\beta}
\def\d{\delta}
\def\e{\epsilon}

\def\g{\gamma}

\def\semi{\bigcirc\kern-1em{s}\;}

\def\del{\partial}
\def\ni{\noindent}

\def\one{{\mathchoice {\rm 1\mskip-4mu l} {\rm 1\mskip-4mu l}
{\rm 1\mskip-4.5mu l} {\rm 1\mskip-5mu l}}}
\def\Q{{\mathchoice
{\setbox0=\hbox{$\displaystyle\rm Q$}\hbox{\raise 0.15\ht0\hbox to0pt
{\kern0.4\wd0\vrule height0.8\ht0\hss}\box0}}
{\setbox0=\hbox{$\textstyle\rm Q$}\hbox{\raise 0.15\ht0\hbox to0pt
{\kern0.4\wd0\vrule height0.8\ht0\hss}\box0}}
{\setbox0=\hbox{$\scriptstyle\rm Q$}\hbox{\raise 0.15\ht0\hbox to0pt
{\kern0.4\wd0\vrule height0.7\ht0\hss}\box0}}
{\setbox0=\hbox{$\scriptscriptstyle\rm Q$}\hbox{\raise 0.15\ht0\hbox to0pt
{\kern0.4\wd0\vrule height0.7\ht0\hss}\box0}}}}
\def\C{{\mathchoice
{\setbox0=\hbox{$\displaystyle\rm C$}\hbox{\hbox to0pt
{\kern0.4\wd0\vrule height0.9\ht0\hss}\box0}}
{\setbox0=\hbox{$\textstyle\rm C$}\hbox{\hbox to0pt
{\kern0.4\wd0\vrule height0.9\ht0\hss}\box0}}
{\setbox0=\hbox{$\scriptstyle\rm C$}\hbox{\hbox to0pt
{\kern0.4\wd0\vrule height0.9\ht0\hss}\box0}}
{\setbox0=\hbox{$\scriptscriptstyle\rm C$}\hbox{\hbox to0pt
{\kern0.4\wd0\vrule height0.9\ht0\hss}\box0}}}}

\font\fivesans=cmss10 at 4.61pt
\font\sevensans=cmss10 at 6.81pt
\font\tensans=cmss10
\newfam\sansfam
\textfont\sansfam=\tensans\scriptfont\sansfam=\sevensans\scriptscriptfont
\sansfam=\fivesans
\def\sans{\fam\sansfam\tensans}
\def\Z{{\mathchoice
{\hbox{$\sans\textstyle Z\kern-0.4em Z$}}
{\hbox{$\sans\textstyle Z\kern-0.4em Z$}}
{\hbox{$\sans\scriptstyle Z\kern-0.3em Z$}}
{\hbox{$\sans\scriptscriptstyle Z\kern-0.2em Z$}}}}

\newcount\foot
\foot=1
\def\note#1{\footnote{${}^{\number\foot}$}{\ftn #1}\advance\foot by 1}

\def\frac#1#2{{#1\over #2}}
\def\text#1{\quad{\hbox{#1}}\quad}

\font\ch=cmbx12 scaled\magstephalf
\font\ftn=cmr8 scaled\magstephalf

\font\it=cmti10 scaled\magstephalf

\font\titch=cmbx12 scaled\magstep2
\font\titname=cmr10 scaled\magstep2
\font\titit=cmti10 scaled\magstep1
\font\titbf=cmbx10 scaled\magstep2

\nopagenumbers

\line{\hfil DFF 228/05/95}
\line{\hfil May 27, 1995}
\vskip2.2cm
\centerline{\titch THE VOLUME OPERATOR IN }
\vskip.5cm
\centerline{\titch DISCRETIZED QUANTUM GRAVITY}
\vskip1.7cm
\centerline{\titname R. Loll\note{Supported by the European Human
Capital and Mobility program on ``Constrained Dynamical Systems"}}
\vskip.5cm
\centerline{\titit Sezione INFN di Firenze}
\vskip.2cm
\centerline{\titit Largo E. Fermi 2}
\vskip.2cm
\centerline{\titit I-50125 Firenze, Italy}

\vskip3.5cm
\centerline{\titbf Abstract}
\vskip0.7cm
We investigate the spectral properties of the
volume operator in quantum gravity in the framework of
a previously introduced lattice discretization. The presence of a
well-defined scalar product in this approach permits us to make
definite statements about the hermiticity of quantum operators.
We find that the spectrum of the volume operator is discrete, but
that the nature of its eigenstates differs from that found in an
earlier continuum treatment.

\vfill\eject
\footline={\hss\tenrm\folio\hss}
\pageno=1


\line{\ch 1 Introduction\hfil}

One of the most active branches of research into the quantization
of 3+1-dimensional gravity of the last few years has been the
canonical, operator-based framework of the so-called loop approach.
It is non-perturbative in the sense that it is not a priori
restricted to the study of geometries close to flat Minkowski
space. Its basic variables are (non-local) generalized Wilson
loops of the $SL(2,\C)$-valued Ashtekar connection.
Also in the quantum theory the state space and operators are
labelled by (equivalence classes of) closed curves in three-space,
which has led to considerable progress in solving the quantum
constraints of the theory. The first, formal solutions to all of
the constraints, including the Wheeler-DeWitt equation, were
found in this loop formulation [1].

Although since then many
of the mathematical ingredients of loop representations have been
scrutinized and better understood (see, for example, [2]), one is
still lacking a rigorous
control over the regularization procedure necessary for obtaining
a well-defined quantum Hamiltonian. One difficulty is the
absence of a natural background metric in the ``fully
diffeomorphism-invariant phase" of the theory. Secondly, since the
basic variables are non-local, the definition of the quantum
Hamiltonian $\hat H$ involves usually a shrinking of loop operators
to points, which arguably is a rather ill-defined process.
These problems, and the absence of a well-defined scalar product
in the quantum representation, have
hampered progress toward a better understanding of the ``solutions
to all
the constraints" and of observables (which in the pure gravity theory
are those operators commuting with the quantum Hamiltonian).

In a recent paper [3], we have proposed an alternative regularization
for the loop approach, that does not involve a point-splitting for
the definition of the Hamiltonian constraint. It is a lattice
regularization of the type used in quantum chromodynamics [4], but
with two important differences.
Firstly, the lattice is considered as purely
topological, and therefore the basic Wilson loop variables of the
theory (with support on the links of the lattice) are both manifestly
gauge- and spatial diffeomorphism-invariant. Secondly, since the
``gauge group" $SL(2,\C)$ is non-compact, we do not use the Haar
measure to define the inner product, but a suitably defined measure
on holomorphic $SL(2,\C)$-valued wave functions, with respect to
which the norm of holomorphic Wilson loop states is finite.
Thus one may think of the construction as a finite approximation of
the usual loop representation, where the support of loops has been
restricted to a fixed cubic, topological lattice, and where the
momenta are smeared out along lattice links. One expects this
approximation to work ever better with increasing lattice size.

The main assets of this lattice model are its computational simplicity
and the existence of a well-defined scalar product.
In a preliminary investigation of the quantum
Hamiltonian we were able to find a large number of solutions to
the Wheeler-DeWitt equation that have finite norm with respect to
this inner product. Furthermore, questions about the selfadjointness
of operators, and in particular observables, are now within reach.
One test of this and other approaches is whether one can define
physically interesting operators that are selfadjoint. (Reality, even
in the classical theory, is a non-trivial requirement in the Ashtekar
formulation, since the basic phase space variables are complex.)
Since our choice of a scalar product corresponds to a choice of
``reality conditions" at the quantum level, and moreover the setting
is fully regularized, it makes sense to associate classical, real
phase space functions with selfadjoint lattice operators.

In the present paper, we will be concerned with the so-called volume
operator, introduced in [5]. Although it is not an observable of the
pure theory, there are arguments suggesting it will become one once
matter has been included. Rovelli and Smolin have presented a partial
computation of its spectrum, based on a certain continuum
regularization [6]. According to their arguments, the spectrum is
both real and discrete, which by them is taken to indicate a
fundamental discreteness of the theory at the Planck scale. This
result is formal in the sense that it postulates the existence of
the quantum operators involved (and the limiting procedure used to
define them), and of a scalar
product that makes the spectrum calculation meaningful.

Given the scalar product of the lattice model, one may in turn ask
whether an analogue of the volume operator can be sensibly
defined and whether its spectrum agrees with that found in the formal
continuum calculation. We will show here that the answer to the first
question is in the affirmative. The lattice regularizes in a natural
way the terms cubic in momenta that appear in the definition of the
volume operator, and its spectrum is again
discrete. However, the nature of the eigenstates (to the extent they
can be compared) disagrees with that found in [6]. In particular, we
find that eigenstates of the volume operator are necessarily
{\it complex} linear combinations of the Wilson loop states.
We will also give a general argument for why trivalent spin network
states are necessarily zero-eigenvectors of the volume operator \note{
The authors of [6] have recently informed me that their derivation of
the spectrum of these states contains a computational error.}.
Finally, we will point out a difficulty that arises in requiring
certain
cubic operators (that are part of the definition of the volume
operator) to be selfadjoint.

In the next section, we recall the construction of the classical
volume function, and define a discretized form of the quantum
volume operator in the holomorphic representation. In Sec.3 we
compute the local action, around a lattice vertex, of the volume
operator on a number of Wilson loop states, and discuss the role
of spin network states as eigenstates of this operator. Finally,
in Sec.4 we compare our method and results with those of the continuum
approach.
\vskip2cm

\line{\ch 2 Defining the volume operator\hfil}

Let us first summarize the main ingredients of the lattice formulation
introduced in [3]. The lattice is a cubic $N\times N\times N$-lattice,
with periodic boundary conditions, i.e. its topology is that of a
three-torus. The basic operators associated with each lattice link $l$
are a holomorphic $SL(2,\C)$-link holonomy $\hat V_A{}^B$ and a
canonical momentum operator $\hat p_i$, with an adjoint index $i$. The
wave functions are elements of $\times_l L^2(SL(2,\C),d\nu_t)^{\cal H}$,
with the product taken over all links. The measure is the heat kernel
measure $d\nu_t$, and the superscript ${\cal H}$ denotes the subset of
holomorphic $L^2$-functions. The basic commutators are

$$
\eqalign{
&[\hat V_A{}^B(n,\hat a),\hat V_C{}^D(m,\hat b)]=0\cr
&[\hat  p_i(n,\hat a),\hat V_A{}^C(m,\hat b)]=
-\frac{i}{2}\,\d_{nm}\d_{\hat a\hat b}\, \tau_{iA}{}^B\hat V_B{}^C\cr
&[\hat p_i(n,\hat a),\hat p_j(m,\hat b)]=
i\, \d_{nm}\d_{\hat a\hat b}\, \e_{ijk}\, \hat p_k,}\eqno(2.1)
$$

\ni where the links are labelled by their initial vertex $n$ and a
positive direction $\hat a$ emanating from it, and $\e_{ijk}$ are the
structure constants of $SU(2)$. In terms of an explicit
parametrization by four complex parameters $\a_i$, $i=0\dots 3$,
$\sum_i\a_i^2 =1$, the operators for a single link $(n,\hat a)$
are given by

$$
\eqalign{
&\hat V_A{}^B=\left( \matrix{\a_0+i\a_1&\a_2+i\a_3\cr
    -\a_2+i\a_3&\a_0-i\a_1}\right)\cr
&\hat p_1=\frac{i}{2} (\a_1\del_0-\a_0\del_1+\a_3\del_2-\a_2\del_3)\cr
&\hat p_2=\frac{i}{2} (\a_2\del_0-\a_3\del_1-\a_0\del_2+\a_1\del_3)\cr
&\hat p_3=\frac{i}{2} (\a_3\del_0+\a_2\del_1-\a_1\del_2-\a_0\del_3).}
\eqno(2.2)
$$

Examples of $SL(2,\C)$-invariant states are given by the Wilson
loops (i.e. the traces of link holonomies)
$Tr\,V(\g)=Tr\,V(l_1)V(l_2)\dots V(l_n)$, where $\g=l_1\circ
l_2\circ\dots l_n$ is a closed lattice loop.
Recall that we do not have an explicit coordinate expression for
the heat kernel measure $d\nu_t$, and therefore must use
the holomorphic transform $C_t:L^2(SU(2),dg)\rightarrow
L^2(SL(2,\C),d\nu_t)^{\cal H}$ and its inverse to compute scalar
products in $L^2(SL(2,\C),d\nu_t)^{\cal H}$. It turns out that the
operators $\hat p_i$ are selfadjoint in the holomorphic representation
(a fact that had been overlooked in [3]),
i.e. they are the holomorphic transforms of the corresponding
selfadjoint differential operators on $L^2(SU(2),dg)$ (whose functional
form coincides with (2.2), with the complex parameters $\a_i$
substituted by real ones). The reason for this is basically that the
$\hat p_i$ map eigenspaces of the Laplacian $\Delta =-4\sum_i \hat
p_i^2$ with fixed eigenvalue $k$ into themselves. Still,
the operators $\hat V_A{}^B$ are {\it not} selfadjoint in the holomorphic
representation, although multiplication by (real) $V_A{}^B$ is a
selfadjoint operation in the $SU(2)$-representation.

The classical expression for the volume of a spatial region $\cal R$
is given by

$$
{\cal V}({\cal R})=\int_{\cal R} d^3x\;\sqrt{\det g}=
\int_{\cal R} d^3x\;\sqrt{\frac{1}{3!} |\e_{abc}\,\e^{ijk} E^a_i
 E^b_j  E^c_k |},\eqno(2.3)
$$

\ni where $E^a_i$ are the momenta of the canonically conjugate
Ashtekar variables $(A_a^i(x), E_i^a(x))$.
In the classical theory, $\cal V$ is of course a real quantity.
A natural lattice discretization of the $\det g$-term is

$$
D(n):=\e_{abc}\,\e^{ijk} p_i(n,\hat a)
p_j(n,\hat b) p_k(n, \hat c),\eqno(2.4)
$$

\ni which in the continuum limit $a\rightarrow 0$ with respect to an
arbitrary lattice spacing $a$ goes over to
$a^6 \e_{abc}\e^{ijk} E^a_i  E^b_j  E^c_k+O (a^7)$.
We may therefore take

$$
{\cal V}_{\rm latt}=\sum_{n\in{\cal R}} \sqrt{
|\e_{abc}\,\e^{ijk}\, p_i(n,\hat a) p_j(n,\hat b)
p_k(n, \hat c)|}\eqno(2.5)
$$

\ni as the lattice analogue of (2.3). The translation of this expression
to the quantum theory is a priori not well defined, because of the
presence of both the modulus and the square root. However, we are in
the fortunate position that the operators

$$
\hat D(n):=\e_{abc}\,\e^{ijk}\, \hat p_i(n,\hat a)
\hat p_j(n,\hat b) \hat p_k(n, \hat c)\eqno(2.6)
$$

\ni are all self-adjoint. (Note that no operator ordering problem occurs
in the definition of $\hat D(n)$ since it contains an
anti-symmetrization over spatial directions.)
We may therefore go to a basis of
$\times_l L^2(SL(2,\C),d\nu_t)^{\cal H}$ consisting of simultaneous
eigenfunctions of all the $\hat D(n)$ and {\it define} the operator

$$
\hat {\cal V}_{\rm latt}=\sum_n \sqrt{|\hat D(n) |}\eqno(2.7)
$$

\ni through the square roots of the moduli of the eigenvalues of the
$\hat D(n)$ in that basis. Thus, all we have to do to understand the
regularized volume operator is to compute the spectra of the
operators $\hat D(n)$.

As mentioned in the introduction, there already exists a (partial)
spectrum calculation in the continuum [6] we can compare with.
Its authors propose to work in terms of a basis of gauge- and
spatially diffeomorphism-invariant states which is diagonal with
respect to
appropriate continuum analogues of the operators $\hat D(n)$
above (whose construction involves smearing $\det g$ over a
small box and then
letting the box shrink to zero). These states are given by
so-called spin networks, constructed from trivalent (or n-valent)
graphs whose
edges are labelled by irreducible representations of $SU(2)$, and
vertices by $SU(2)$-intertwining operators (see, for
example, [7] and references therein).
They can be thought of as certain (anti)symmetrized, real linear
combinations of multiple
Wilson loops with support on the graph. The difference with our
discrete formulation is that one considers all possible graphs
(with all possible labellings), whereas we keep the lattice fixed, and
therefore the total number of degrees of freedom finite.
Still, also the lattice approach allows for a similar construction
of gauge-invariant states. Finding an efficient labelling for such
states is a well-known problem in lattice gauge theory, and various
methods have been used (see, for example, [8]), among them
constructions reminiscent of the spin networks, generalized
to cubic lattices (which have intersections of order higher than
three). A problem that typically occurs is that such explicitly
gauge-invariant bases are overcomplete, and that for doing
computations one needs an efficient way of labelling a complete
subset of independent states. For spin network states of valence
higher than three, one encounters a similar problem. Whether one
basis is better than another is determined by the dynamics of the
basic operators of the theory, and may therefore be completely
different for gravitational and gauge-theoretic applications.

When studying the volume operator it is indeed useful to consider
a representation in which the lattice links are labelled by positive
``occupation numbers", which count the number of (unoriented)
flux lines of
basic spin-$\frac12$ representations on the link (with the links
contracted gauge-invariantly at the vertices). In essence this
happens because the operators $\hat D(n)$ have a particularly
simple structure: when acting on a multiple Wilson loop, they do not
change its support (in terms of the flux line numbers), and only some
finite-dimensional rearrangements occur within the subset of states
that share the same occupation numbers. In the context of the
continuum theory, a similar observation was
already made in [6]. In this respect, the volume operator is much
simpler than the Hamiltonian operator, which (at least on the
lattice) changes the support of a Wilson loop state it acts on [3].

The explicit part of the continuum spectrum calculation in [6]
was made for trivalent spin networks.
Although general gauge-invariant lattice states contain six-valent
intersections, one can easily construct states that are only trivalent
by assigning the occupation number zero to an appropriate subset of
lattice links. (By contrast, in the continuum picture, assigning zero
occupation number to an edge can be interpreted as creating a new,
smaller graph.)
The question therefore arises whether spin networks constructed from
such trivalent states are also eigenstates in the lattice formulation.
To answer it, it is
sufficient to study the action of the operators $\hat D(n)$, as
explained above. This will be the subject of the next section.

\vskip2cm

\line{\ch 3 Computation of the spectrum\hfil}

We will now present the results of the spectral computation
for small occupation numbers around a single vertex $n$, which will
be sufficient to illustrate our
point; a complete construction will appear elsewhere [9].
It turns out that the spectrum of $\hat D(n)$ is discrete. This was
not clear a priori, since the group $SL(2,\C)$ is non-compact; it is
a consequence of our choice of a scalar product. We will not
speculate here whether this discreteness is of a fundamental
nature or only an artifact of the regularization, that will disappear
in an appropriately taken continuum limit.

\epsffile{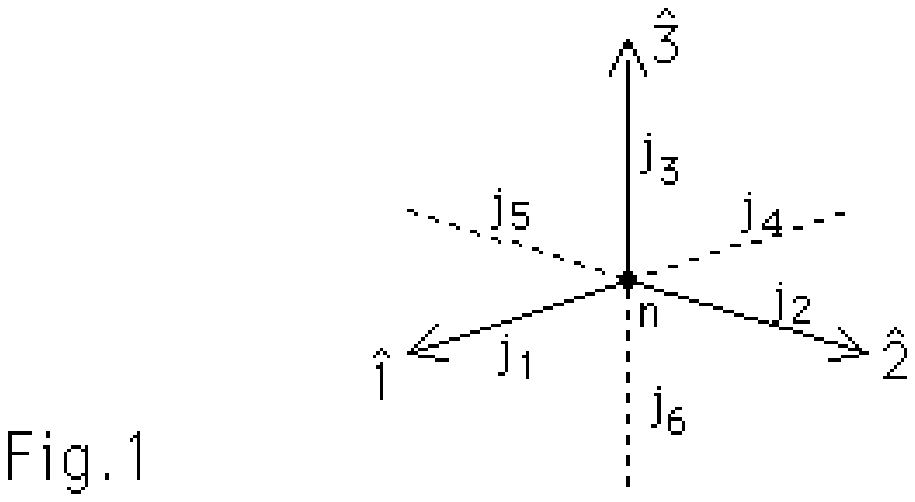}

Fig.1 illustrates the labelling of the link directions meeting at a
vertex $n$. We will be interested in the behaviour of gauge-invariant
states under the action of $\hat D(n)$. One ingredient in the
labelling of such a state is a 6-tuple $\vec j$ of integers $j_i\geq 0$
giving the occupation numbers $(j_1,\dots,j_6)$ of the links
$((n,\hat 1),(n,\hat 2),(n,\hat 3);(n,-\hat 1),(n,-\hat 2),(n,-\hat 3))
\equiv ((n,\hat 1),(n,\hat 2),(n,\hat 3);
(n-\hat 1,\hat 1),(n-\hat 2,\hat 2),
(n-\hat 3,\hat 3))$ intersecting at $n$. We will call $j:=\sum_{i=1}^6
j_i$ the {\it order} of a state (at $n$), which is an even integer.
What remains to be
specified is the way the $j$ flux lines are joined pairwise at $n$
to ensure gauge-invariance. By convention we allow a flux line coming
in from the positive $\hat 1$-direction, say, to be joined only to a
flux line from one of the other five links, and not from the same
link (i.e. we forbid ``retracings"). This leads to a constraint on
the occupation numbers: any $j_i$ has to be equal to or smaller than
the sum of the remaining $j_k$, for example, $j_6\leq\sum_{i=1}^5
j_i$.

Given $\vec j$, the number of possible different contractions of flux
lines at $n$ is finite. Not all of them will lead to linearly
independent Wilson loop states once the flux line configuration around
$n$ is extended to a gauge-invariant state of lattice loops. To
understand this, think of a fixed (but arbitrary) such extension of
the flux line configuration, so as to obtain a set of closed lattice
curves. They may be thought of as a set $\g_1$, $\g_2$,
$\dots$, $\g_k$ of lattice loops based at $n$, and the corresponding
multiple Wilson loop is the state $\Psi=Tr\,V(\g_1)\,Tr\,V(\g_2)\dots
Tr\,V(\g_k)$.
Now, different contractions of the flux lines at $n$ will lead to
different Wilson loop states (with the same support), which in general
are related by so-called Mandelstam constraints. For example, for the
case of three lattice loops meeting at $n$, one has the following
identity [10]:

$$
\eqalign{
Tr\,V(\g_1)\,Tr\,V&(\g_2)\,Tr\,V(\g_3)=
Tr\,V(\g_1)\,Tr\,V(\g_2\circ\g_3)+
Tr\,V(\g_2)\,Tr\,V(\g_1\circ\g_3)+\cr
&Tr\,V(\g_3)\,Tr\,V(\g_1\circ\g_2)-
Tr\,V(\g_1\circ\g_2\circ\g_3)-
Tr\,V(\g_2\circ\g_1\circ\g_3)}\eqno(3.1).
$$

For the special case of a trivalent graph, Rovelli and Smolin have given
a prescription for associating with each labelling of flux lines a
{\it unique} quantum state, obtained by appropriately
(anti)symmetrizing
over all possible Wilson loop states sharing the same flux labels [6]:
calling temporarily $j(l)$ the occupation number of a link $l$, the
number of different multiloops one can associate with a given flux
line labelling -- by permuting the way individual flux lines are joined
at vertices -- is $\prod_l j(l)!$, where the product is taken over all
lattice links. The spin network state is then obtained by adding
the corresponding $\prod_l j(l)!$ Wilson loop states, with the
weight $(-1)^{(p+n)}$, where $p$ is the parity of the flux line
permutation and $n$ the number of closed loops in the multiloop.

The case of trivalent intersections turns out to be particularly
simple, since there is only one way of contracting the
(anti)symmetrized flux line configurations at each vertex.
Translated to the lattice, and according
to the reasoning at the end of the previous section, this means that
$\hat D(n)\Psi = d\,\Psi$ for any trivalent spin network state $\Psi$
(i.e. a gauge-invariant state on the lattice with at most
trivalent intersections -- at each vertex $n$, at least three of the
flux line labels of the adjacent links are zero). That is, $\Psi$ is
necessarily an eigenstate of $\hat D(n)$, with eigenvalue $d$ (where
of course $d$ depends on $\Psi$).

Let us now compute the action of the operator $\hat D(n)$ on some
trivalent lattice spin networks. Recall that the momenta in
$\hat D(n)$,
(2.6), act non-trivially on the lattice links emanating from the
vertex $n$ in a positive direction. The simplest type of configuration
is of order 4 and has $\vec j=(2,1,1;0,0,0)$, Fig.2a. There are
two possible permutations of the flux lines, illustrated in Fig.2b.
The corresponding spin network state $\Psi$ is the sum of the
two, $\Psi=\psi_1 +\psi_2$. One finds $\hat D(n)\psi_1=0$ and
$\hat D(n)\psi_2=0$, and therefore $\Psi$ is an eigenvector with
eigenvalue zero.

\epsffile{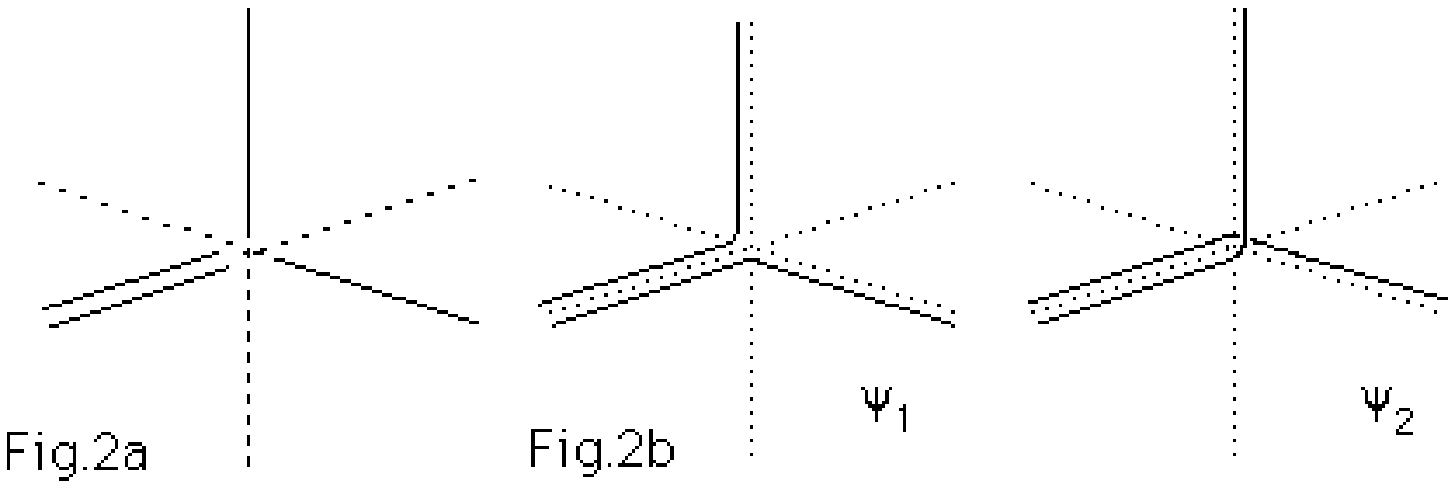}

At order $j=6$ there are two admissible flux line labellings (up
to a permutation of link labels). The
first one is $\vec j =(2,2,2;0,0,0)$, where there are
$2!2!2!=8$ flux line permutations, leading to Wilson loops states
$\psi_i$, $i=1,\dots,8$. One computes $\hat D(n)\psi_i=0$, $\forall
i$, and the spin network state, which is again the weighted sum of the
$\psi_i$, is a zero-eigenvector of $\hat D(n)$. Similarly, for the spin
network state $\Psi$ associated with $\vec j=(3,2,1;0,0,0)$,
one finds $\hat D(n)\Psi=0$.

\epsffile{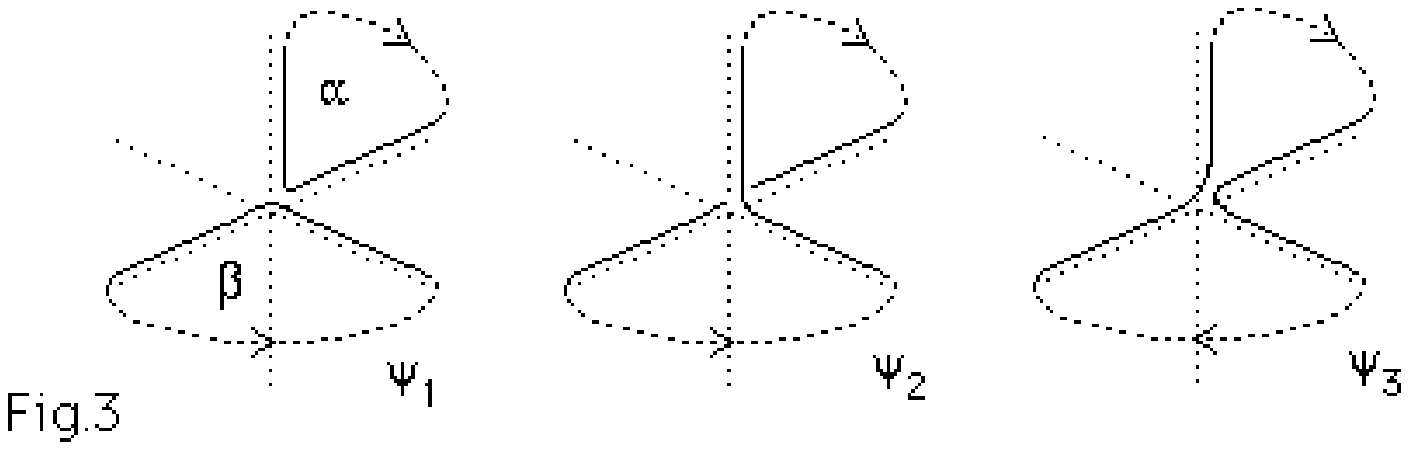}

We will explain shortly why indeed $\hat D(n)\Psi=0$ for {\it any}
trivalent spin network $\Psi$. Before doing so, let us look at
a couple of examples that lead to non-trivial eigenstates
of $\hat D(n)$. First, consider $\vec j=(1,1,1;1,0,0)$. The
three possible contractions at $n$ are illustrated in Fig.3, where
the dotted lines with arrows
denote an arbitrary extension by other lattice
links. The three possible Wilson loop states are $\psi_1 =Tr\,V(\a)
V(\b)$, $\psi_2=Tr\,V(\a\circ\b)$ and $\psi_3=Tr\,V(a\circ\b^{-1})$,
where we have assigned a definite orientation to the composite
loops $\a$ and $\b$. (Following [11], one may also write
the Mandelstam constraints in a way that is independent of the
loop extensions.)
The $\psi_i$ are already spin networks in the sense that
there are no flux line
permutations to be taken into account.
The action of $\hat D(n)$ yields

$$
\hat D(n)\left( \matrix{\psi_1\cr \psi_2\cr \psi_3}\right) =
\frac{3i}{2}\left( \matrix{0&-1&1\cr 2&-1&0\cr -2&0&1}\right)
\left(\matrix{\psi_1\cr\psi_2\cr\psi_3}\right),\eqno(3.2)
$$

\ni and its eigenvectors are easily computed,

$$
\eqalign{
&\hat D(n)\Psi =0,\quad \Psi=\psi_1-\psi_2-\psi_3\cr
&\hat D(n)\,\Psi =-\frac{3\sqrt{3}}{2}\Psi,\quad \Psi=
\psi_1-\frac14 (1-i\sqrt{3})\psi_2-\frac14 (1+i\sqrt{3})\psi_3\cr
&\hat D(n)\,\Psi =\frac{3\sqrt{3}}{2}\Psi,\quad \Psi=
\psi_1-\frac14 (1+i\sqrt{3})\psi_2-\frac14 (1-i\sqrt{3})\psi_3.}
\eqno(3.3)
$$

The presence of the zero-eigenvector is not surprising, since the
three states $\psi_i$ are not independent from the outset, but
rather obey the Mandelstam constraint $\psi_1 -\psi_2 -\psi_3=0$.
There are only two linearly independent spin networks,
and we could have removed the redundancy before looking for
eigenstates of $\hat D(n)$.

A non-trivial example of order 6 is given by $\vec j=(2,1,1;1,1,0)$.
We only sketch the result: there are 12 different configurations
to start with, from contracting the flux lines at $n$. Symmetrization
with respect to the two flux lines in the $\hat 1$-direction leaves us
with 6 spin network states. After using the
Mandelstam constraints [10],
only three linearly independent spin network states remain.
Diagonalizing the action of $\hat D(n)$ on those states,
one finds the three eigenvalues,
$0$, $-3\sqrt{2}$ and $3\sqrt{2}$.

Finally, let us analyze the case $\vec j=(3,1,1;1,0,0)$. There are
6 different ways of contracting the flux lines. After symmetrizing
over the six permutations of the flux lines in the
$\hat 1$-direction, a single spin network state $\Psi$ is obtained,
and one finds $\hat D(n)\Psi=0$. Interestingly, if one looks at
the Wilson loop states before the symmetrization, one can form two
linear combinations that are non-trivial eigenstates of $\hat D(n)$,
with eigenvalues $\pm \frac32 \sqrt{3}$. Thus it seems as if one were
losing information about the gauge-invariant sector of the Hilbert
space by looking only at the spin network states.
However, this is presumably not the case, since the information
may be contained in other spin networks, associated with a different
labelling of flux lines.

The above examples show that it is possible to find eigenstates
of spin network-type whose eigenvalues are non-zero. However, in
all cases where at a vertex $n$ one can construct only a {\it single}
spin network state (which therefore must be an eigenstate of
$\hat D(n)$), its eigenvalue necessarily vanishes.
As explained earlier, all trivalent vertices are of this type.
This happens
because the momenta $\hat p_i$ in our representation are
represented selfadjointly and according to (2.2) contain each a
factor of $i$ (or rather $i\hbar$), so that also $\hat D(n)$ is
proportional to $i$. It is however easy to see that $\frac1i \hat
D(n)$ maps a Wilson loop state $Tr\,V(\g)$ into a {\it real}
linear combination of such states. Therefore, if we have a
spin network state $\Psi$ (that by construction is a real linear
combination of Wilson loop states), its eigenvalue equation is
$\hat D(n)\Psi=d\,\Psi$, with {\it imaginary} $d$. On the other
hand, $\hat D(n)$ is a selfadjoint operator and its eigenvalues
are real. We hence conclude that necessarily $d=0$.
Non-zero eigenvalues can only occur for intersections that are
higher than trivalent, and the corresponding eigenvectors are
always {\it complex} linear combinations of Wilson loop states,
as illustrated by (3.3).

\vskip2cm

\line{\ch 4 Discussion and conclusions\hfil}

The calculations of the previous section took place around a single
vertex $n$, but can be generalized immediately
to lattice regions $\cal R$ containing several or even all of the
lattice vertices, to obtain eigenstates of the
volume operator $\hat {\cal V}_{\rm latt}({\cal R})$.
Although its spectrum is obviously discrete, we have found that
all quantum spin network states corresponding to trivalent graphs
are eigenstates with eigenvalue zero. This disagrees with the
continuum computation of the trivalent sector
reported by Rovelli and Smolin [6], where a non-vanishing
spectrum was found. It is therefore important to understand how
the two regularization methods differ.

The presence of factors of $i$ in the definition of our momentum
operators $\hat p_i$ can be traced back to the canonical
commutators of the continuum Yang-Mills theory,
$[\hat A_a^i (x),\hat E_j^b (y)]=i\,\d_a^b \d_j^i \d^3 (x-y)$,
whose lattice analogues in the holomorphic representation
are given by (2.1). However, we strictly speaking should be
quantizing the classical Poisson brackets
$\{ A_a^i (x), E_j^b (y)\}=i\,\d_a^b \d_j^i \d^3 (x-y)$
of the canonical Ashtekar variables [12], leading to canonical
commutators

$$
[\hat A_a^i (x),\hat E_j^b (y)]=\d_a^b \d_j^i \d^3 (x-y)
\eqno(4.1)
$$

\ni {\it without} a factor $i$. In [3], we quantized the commutators
{\it with} the factor $i$, to facilitate comparison with the usual
formalism of Hamiltonian lattice gauge theory. This was done
with the understanding that the quantum commutators with and
without $i$ can in a straightforward way be related by multiplying the
canonical momenta by $i$. For instance, for the case of our
lattice variables, defining new momenta $\hat p_i':=i\,\hat p_i$
leads to a version of the basic commutator algebra (2.1) without
any factors of $i$ on the right-hand sides. In fact, a representation
of the form $\hat E=\del/\del A$ for
the canonical momentum operators has been used both in formal
continuum formulations [1,13] and in previous lattice approaches to
gravity [14].
Would the substitution $\hat p_i\rightarrow \hat p_i'$ change the
results on the spectrum of the volume operator obtained in Sec.3?
Obviously not: the operators $\hat p_i'$ and the corresponding
composite
operators $\hat D(n)'$ would become anti-hermitian, and their
spectra purely imaginary. Otherwise, the spectra would of course
remain discrete, and trivalent spin network states would still
be eigenstates with zero eigenvalues, leaving our main results
unaffected.

However, insisting that -- for whatever physical reasons --
the operators $\hat D(n)$ be selfadjoint,
one seems forced to adopt a representation
like $\hat E=\frac1i \,\del/\del A+\dots$, where the dots stand for
possible divergence terms. This is at least true if the
representation is defined on a Hilbert space of states on
${\cal A}^{SL(2,\C)}/{\cal G}^{SL(2,\C)}$ of connections modulo
gauge, or an appropriate generalization thereof (our lattice Hilbert
space is a discretized version of this space).
This is an illustration of the well-known fact that
requiring certain physical operators to be selfadjoint leads
to restrictions on the possible quantum representations. One
possible set of selfadjointness conditions for continuum gravity
in the connection representation is to demand that
$\hat E_i^a\hat E^{bi}$ and $[\hat H,\hat E_i^a\hat E^{bi}]$
to be selfadjoint, which are the quantum counterparts of the
reality conditions on the classical spatial three-metric $q_{ab}$,
$\det q\, q^{ab}=E^a_i E^{bi}$, and its evolution
$\{H,E_i^a E^{bi}\}$, where $H$ denotes the Hamiltonian.
Normally these ``quantum reality conditions" are ill-defined,
because they contain products of quantum operators at the same point.
However, the basic variables $(V_A{}^B,p_i)$ of the lattice
formulation are already regularized appropriately (recall also
that the spatial diffeomorphisms have already been factored out),
and it is therefore well-defined to require $\hat p_i(n,\hat a)
\hat p^i (n,\hat b)$ to be selfadjoint. This is compatible with
both selfadjoint and anti-selfadjoint momenta $\hat p_i$.
Demanding in addition the selfadjointness of
$\e_{abc}\,\e^{ijk}\, \hat p_i(n,\hat a)
\hat p_j(n,\hat b) \hat p_k(n, \hat c)$ excludes the possibility
of having anti-selfadjoint momenta with $\hat p_i^\dagger =-\hat p_i$.
Alternatively, if one is only interested in the selfadjointness of
the volume operator $\hat {\cal V}_{\rm latt}({\cal R})$ (and not of
the $\hat D(n)$), one may leave the momenta $\hat p_i$
anti-selfadjoint,
in which case the modulus in (2.7) takes care of turning the
eigenvalues of $\hat {\cal V}_{\rm latt}({\cal R})$ into positive real
numbers.

In the continuum treatment of [6], the definition of the volume
operator involves a quantized version of the generalized Wilson
loop variable with three momentum insertions (at loop parameters
$s$, $t$ and $r$),

$$
T^{abc}[\a](s,t,r)=Tr\, E^a(\a(s))V_\a(s,t) E^b(\a(t))V_\a(t,r)
E^c(\a(r))V_\a(r,s),\eqno(4.2)
$$

\ni in the limit as the loop argument $\a$ shrinks to a point.
Note that it is only in this limit that the classical variable
(4.2) becomes real, since the holonomies $V_\a$ depend on the
{\it complex} Ashtekar connections $A_a$ -- only for the
``point loop" $\a$, one has a real holonomy $V_\a=\one$.
This problem does not arise if one interprets the calculations
as taking place within the Euclidean theory (with the corresponding
scalar product), where $V_\a$ is real
from the outset.

In the corresponding limiting procedure in the quantum theory
one lets an auxiliary length variable $L$ (measuring the edge
length of a small box) go to zero and defines the volume
operator as $\hat {\cal V}({\cal R}):=\lim_{L\rightarrow 0}
\hat {\cal V}_L ({\cal R})$ [6]. Unfortunately, as already noted
there, it is difficult to make rigorous sense of this limit
since for any $L\not= 0$, the operator $\hat {\cal V}_L$
is not even formally selfadjoint.
As long as $L$ is finite, acting with
$\hat {\cal V}_L$ on a Wilson loop changes its support, so
there is no obvious basis of loop
states for which it is diagonal, and one cannot make sense of
the square root operation.
Thus, even if this construction does in the end lead to a finite
operator, the status of the regularization procedure is still
unclear.
In any case, although the (left) spectrum
of $\hat {\cal V}( {\cal R})$ is allegedly real, no scalar
product was specified in [6], with respect to which its selfadjointness
or otherwise could be established.

The differences between the continuum and lattice regularizations
make a direct comparison of the spectral computations difficult.
Also it is not a priori clear to what extent they should agree, given
that no continuum limit has yet been performed in the lattice
formulation. Even if eventually an agreement on the vanishing of
$\hat {\cal V}_{\rm latt}$ on trivalent states can be reached, it
does not automatically follow that the non-zero parts of the spectra
will coincide in both formalisms.
The fact that in our approach the trivalent spin network
states all ``have zero volume" may be taken as an indication that they
are degenerate from a physical point of view. In any case, it
makes clear that considering trivalent states alone is not
enough. --
We are currently addressing the question of how to construct a
non-overcomplete basis of holomorphic lattice states in terms
of which the Hamiltonian and volume operators assume a simple form,
which hopefully will lead to a complete spectral analysis of
$\hat {\cal V}_{\rm latt}$ [9].

We also have pointed out that the requirement of having
the $\e_{abc}\e^{ijk}\hat E^a_i \hat E^b_j \hat E^c_k$, or some
suitably smeared versions thereof, represented by selfadjoint
operators suggests a quantum
representation $\hat E^a_i=\frac1i\, \del/\del A_a^i$ for the
momenta, if the Hilbert
space is a space of connections, as is usually the case.
This is at odds with the commutation relations (4.1), if one
represents $\hat A$ as the multiplication operator by $A$.
Whether or not this distinction has any physical relevance remains
to be seen; it is an example of how selfadjointness conditions on
operators restrict the choice of possible quantum representations.
\vskip1cm
\ni{\it Acknowledgement.} I would like to thank the participants of
the Warsaw workshop on canonical and quantum
gravity for discussion and comments.

\vskip2cm

\line{\ch References\hfil}

\item{[1]} Rovelli, C. and Smolin, L.: Loop space representation of
  quantum general relativity, {\it Nucl. Phys.} B331 (1990) 80-152

\item{[2]} Ashtekar, A. and Isham, C.J.: Representations of the
holonomy algebras of gravity and non-Abelian gauge theories,
{\it Class. Quant. Grav.} 9 (1992) 1433-67; Ashtekar, A. and
Lewandowski, J.: Representation theory of analytic holonomy
$C^*$-algebras, in {\it Knots and quantum gravity}, ed. J. Baez,
Clarendon Press (Oxford) 1994, 21-61; Ashtekar, A., Lewandowski, J.,
Marolf, D., Mour\~ao, J. and Thiemann, T.: Quantization of
diffeomorphism invariant theories of connections with local
degrees of freedom, {\it preprint} Penn State U., Apr 1995,
e-Print Archive: gr-qc 9504018

\item{[3]} Loll, R.: Non-perturbative solutions for lattice quantum
  gravity, to appear in {\it Nucl. Phys.} B,
  e-Print Archive: gr-qc 9502006

\item{[4]} Kogut, J. and Susskind, L.: Hamiltonian formulation of
  Wilson's lattice gauge theories, {\it Phys. Rev.} D11 (1975)
  395-408; Kogut, J.B.: The lattice gauge theory approach
  to quantum chromodynamics, {\it Rev. Mod. Phys.} 55 (1983) 775-836

\item{[5]} Smolin, L.: Recent developments in nonperturbative
quantum gravity, in {\it Proc. Sant Feliu de Guixols 1991,
Quantum gravity and cosmology}, World Scientific, Singapore, 1992, 3-84

\item{[6]} Rovelli, C. and Smolin, L.: Discreteness of area and
  volume in quantum gravity, {\it Nucl. Phys.} B442 (1995), in press

\item{[7]} Baez, J.B.: Spin network states in gauge theory, {\it
  preprint} UC Riverside, Nov 1994, e-Print Archive: gr-qc 9411007;
  Spin networks in nonperturbative quantum gravity, {\it preprint}
  UC Riverside, Apr 1995, e-Print Archive: gr-qc 9504036

\item{[8]} Furmanski, W. and Kolawa, A.: Yang-Mills vacuum: an
  attempt at lattice loop calculus, {\it Nucl. Phys.} B291 (1987)
  594-628

\item{[9]} Loll, R.: {\it preprint} INFN Firenze, in preparation

\item{[10]} Loll, R.: Independent SU(2)-loop variables and the reduced
  configuration space of SU(2)-lattice gauge theory, {\it Nucl. Phys.}
  B368 (1992) 121-42; Yang-Mills theory without Mandelstam
  constraints, {\it Nucl. Phys.} B400 (1993) 126-44

\item{[11]} Rovelli, C. and Smolin, L.: Spin networks and quantum
  gravity, {\it preprint} U. Pittsburgh, Apr 1995, e-Print Archive:
  gr-qc 9505006

\item{[12]} Ashtekar, A.: New variables for classical and quantum
  gravity, {\it Phys. Rev. Lett.} 57 (1986) 2244-7; A new
  Hamiltonian formulation of general relativity, {\it Phys.
  Rev.} D36 (1987) 1587-1603;
 {\it Lectures on non-perturbative canonical
  gravity}, World Scientific, Singapore, 1991

\item{[13]} Jacobson, T. and Smolin, L.: Nonperturbative quantum
  geometries, {\it Nucl. Phys.} B299 (1988) 295-345;
Br\"ugmann, B.: Loop representations, in {\it Canonical
  gravity: from classical to quantum}, ed. J. Ehlers and
  H. Friedrich, Lecture Notes in Physics 434, Springer, Berlin,
  1994

\item{[14]} Renteln, P. and Smolin, L.: A lattice approach to spinorial
  quantum gravity, {\it Class. Quant. Grav.} 6 (1989) 275-94;
Renteln, P.: Some results of SU(2) spinorial lattice
  gravity, {\it Class. Quant. Grav.} 7 (1990) 493-502

\end